\documentclass[a4paper,12pt]{article}
\usepackage[fleqn]{amsmath}
\usepackage[english]{babel}
\usepackage{amssymb}
\usepackage{harvard}
\usepackage{graphicx}
\usepackage{epsfig}      
\usepackage{graphics}
\usepackage{overpic}
\usepackage{rotating}
\usepackage{multirow}
\usepackage{subcaption}
\usepackage{bbm}
\usepackage{url}
\usepackage{hyperref}
\usepackage{tikz}
\usepackage{comment}
\usepackage{xcolor}
\usepackage[export]{adjustbox}
\usepackage{algorithm}
\usepackage{algpseudocode}
\usepackage{framed}
\usepackage{makecell}
\numberwithin{equation}{section}
\usepackage[titletoc,toc,title,page]{appendix}
\usepackage[flushleft]{threeparttable}

\def\bee{\begin{eqnarray}}
\def\eee{\end{eqnarray}}
\def\beee{\begin{eqnarray*}}
\def\eeee{\end{eqnarray*}}

\def\be{\begin{equation}}
\def\ee{\end{equation}}

\newcommand*{\vcenteredhbox}[1]{\begin{tabular}{@{}c@{}}#1\end{tabular}}

\begin{document}

  \citationstyle{dcu}
  \bibliographystyle{dcu}
  \citationmode{default}
\begin{center}
\begin{Large}
\bf Visualizing Topic Uncertainty in Topic Modelling\footnote{Financial support from the German Research Foundation (DFG) (WI 2024/8-1) and the National Science Centre (NCN) (Beethoven Classic 3: UMO-2018/31/G/HS4/00869) for the project TEXTMOD is gratefully acknowledged. The project also benefited from cooperation within HiTEC Cost Action CA 21163. I am indebeted to Viktoriia Naboka for helpful comments on a preliminary version of this paper.}\\[30pt]
\end{Large}

\begin{small}



{Peter Winker}\\
Justus Liebig University Giessen\\
Licher Strasse 64, 35394 Giessen, Germany\\
email: peter.winker@wirtschaft.uni-giessen.de 
\end{small}

\end{center}

\noindent {\bf Abstract}
Word clouds became a standard tool for presenting results of natural language processing methods such as topic modelling. They exhibit most important words, where word size is often chosen proportional to the relevance of words within a topic. In the latent Dirichlet allocation (LDA) model, word clouds are graphical presentations of a vector of weights for words within a topic. These vectors are the result of a statistical procedure based on a specific corpus. Therefore, they are subject to uncertainty coming from different sources as sample selection, random components in the optimization algorithm, or parameter settings. A novel approach for presenting word clouds including information on such types of uncertainty is introduced and illustrated with an application of the LDA model to conference abstracts.\\


\section{Introduction}
One task in natural language processing consists in identifying themes or topics of documents, which might be helpful both for finding documents of interest for a reader or -- at an aggregate level -- for describing the discourse in specific media or over time. For example, \citeasnoun{Luedering2016} search for topics related to economic key indicators and analyze to what extent the importance of these topics in academic literature co-evolves with the actual indicator, \citeasnoun{Lenz2020} describe diffusion of innovations based on topics in news articles, \citeasnoun{Ellingsen2022} use the importance of specific topics as additional input for business forecasting, and \citeasnoun{Savin2022} analyze the development of the topics of nationwide phone-ins with Vladimir Putin public over time and their link to economic indicators. 

For all applications of this type, it is necessary to identify the relevant topics for a given corpus of documents. For this task, many different methods have been proposed, which all have to deal with the high dimensionality of textual data and, consequently, a high computational complexity. Therefore, the algorithms typically can provide only approximate solutions to the problem. Further uncertainty is introduced by setting parameters for the methods and, as a central step, by selecting the number of topics \cite{Bystrov2022b}. As the focus of this paper is not on extending the methods for identifying topics, only a standard approach, the latent Dirichlet allocation (LDA) will be used for demonstrating the novel approach for visualizing topic uncertainty. Nevertheless, the approach might be used universally in situations when topics are obtained from textual data by a mechanism involving random components or meta parameters selected by the researcher, which might be linked to sample selection, pre-processing of data, model assumptions or the estimation algorithm. 

One main result of a topic modelling approach such as LDA is a set of topics, where each topic is a vector of weights over all words in the vocabulary of the underlying corpus (after some pre-processing steps). Given that the vocabulary typically comprises a large number of words, it appears sensible focusing on the most relevant words within a topic, i.e., the words having the highest probability of showing up in a document on this topic. While it is straightforward to generate lists for each topic containing the $l$ most common words and their corresponding weights, this way of presenting the outcome is not very intuitive. Therefore, word clouds became a standard tool for presenting these results. These are figures exhibiting most important words of a topic, where the font size of the word is often chosen at least as a monotonic function of word weights or even proportional to them. 

\begin{figure}[H]
\centering
\vcenteredhbox{\includegraphics[width=0.32\textwidth]{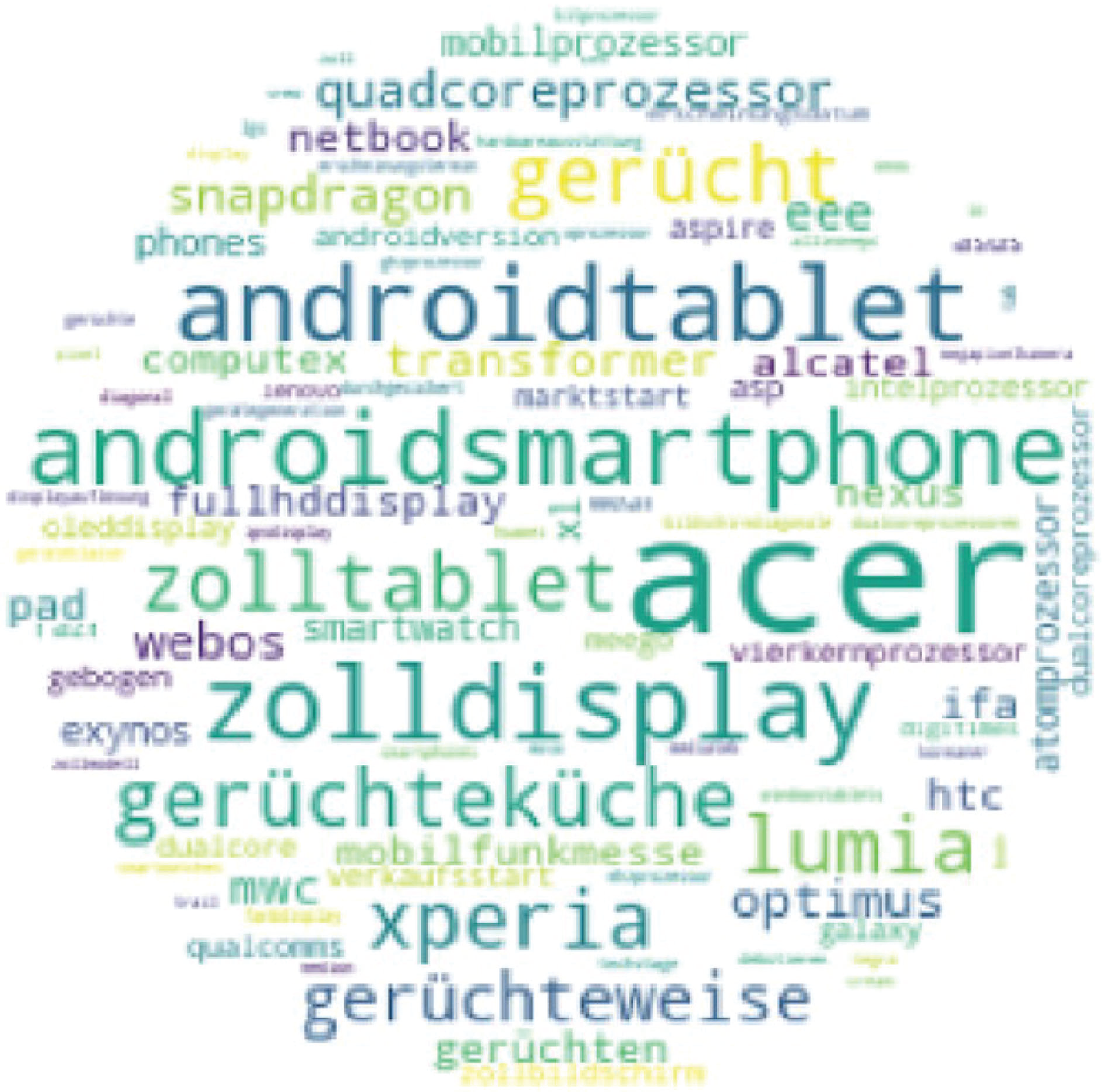}}
\vcenteredhbox{\includegraphics[width=0.32\textwidth]{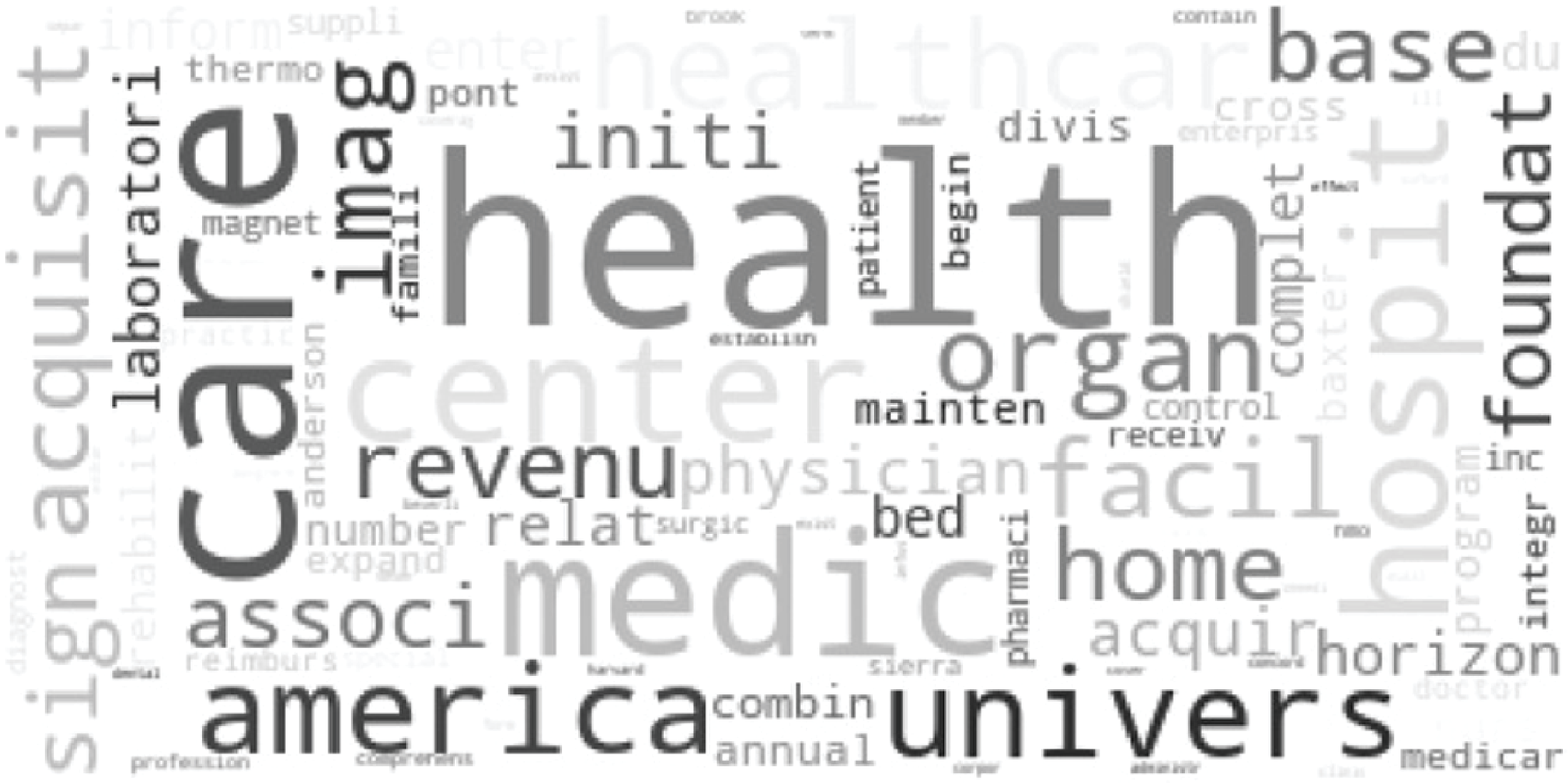}}
\vcenteredhbox{\includegraphics[width=0.32\textwidth]{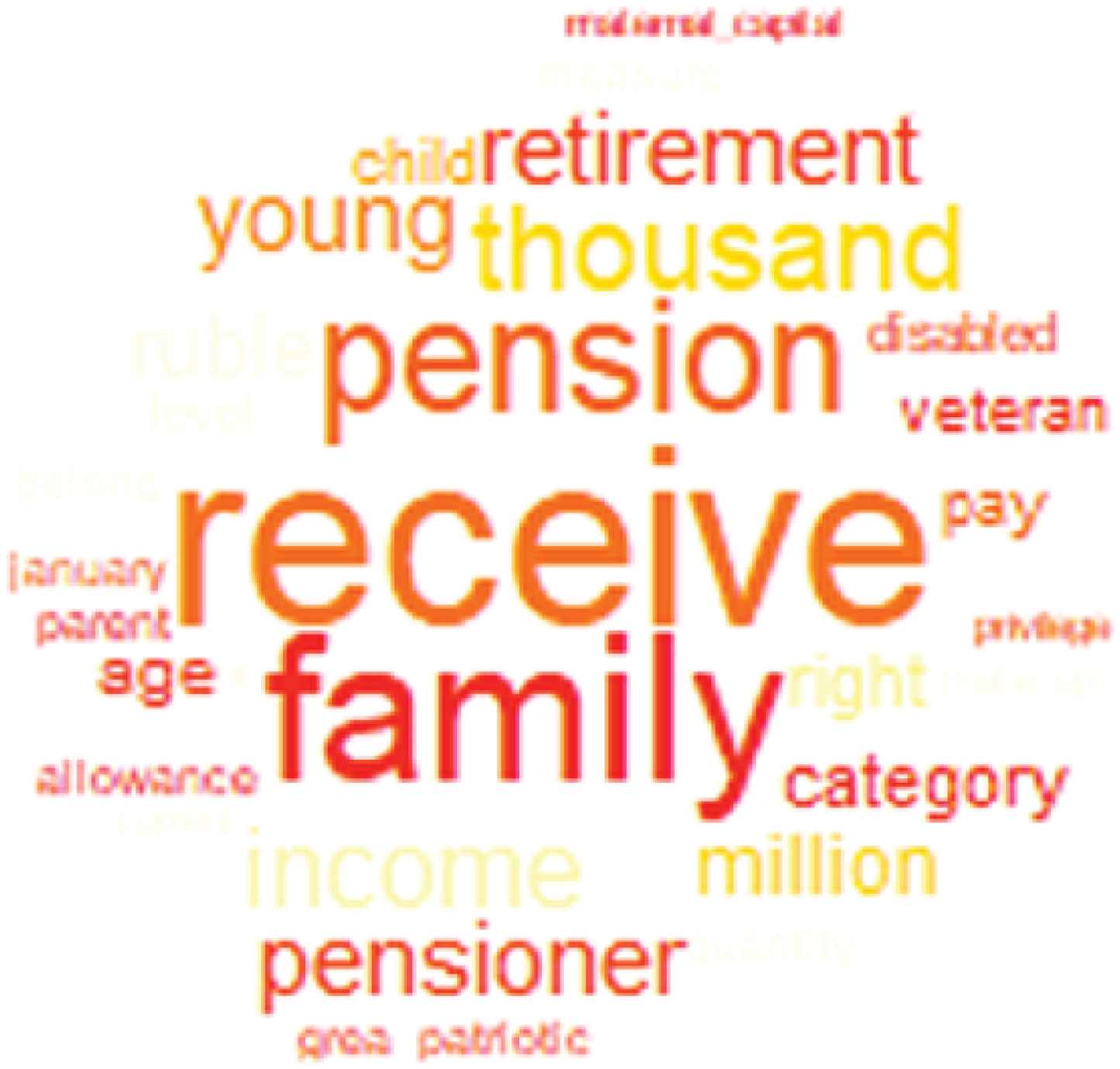}}
\caption{Examples of Word Clouds.}
\label{word_clouds}
\end{figure}

Figure~\ref{word_clouds} shows three examples of such word clouds.\footnote{These figures are taken from \citeasnoun{Lenz2020}, \citeasnoun{Ellingsen2022}, and \citeasnoun{Savin2022}, respectively.} In the original publications, these word clouds and the corresponding topics are also assigned a heading. Just looking at the graphical presentations makes it easy for the reader to agree to these headings, which are "android tablet computers", "health care",  and "social security", respectively.

Thus, the common use of word clouds in the context of topic modelling might be explained by the ease of grasping the content of a topic based on this visualization. It is similar to the ease with which a human reader can interpret a time series plot as compared to studying a list of numbers containing exactly the same information. A major difference from a statistical point of view is that there exist straightforward methods for describing the (statistical) uncertainty related to time series by means of confidence bands \cite{LUTKEPOHL2020} or fan-charts \cite{Britton1998}, while this is not the case for word clouds.\footnote{A related approach in bioinformatics is presented by \citeasnoun{Schneider1990}, where the size of letters correspond to the frequency of aligned patterns in DNA, RNA or protein sequences. However, the limited number of letters in these applications does not allow for a straightforward transfer to word clouds in natural language processing.}  However, also for the generation of word clouds, e.g., based on the LDA model, several sources of uncertainty exist, be it due to the sampling documents into the corpus or random components in the optimization routines used for estimating the models. 

The innovative contribution of this paper is the presentation of a novel approach for presenting uncertainty for word clouds. Thereby, the ease of graphical visualizations is maintained by providing a kind of {\it confidence interval} for each word in a word cloud. The method and its implementation will be described and illustrated with an application of the LDA model to the abstracts of the CFE-CMStatistics conference series 

The remainder of this paper is structured as follows. Section~\ref{measuring} introduces the simulation based method for measuring uncertainty, which has to take into account the lack of identification of the ordering of topics. Section~\ref{graphical} provides the graphical tool for exhibiting this uncertainty including some comments on the computational implementation. An application of the methods to conference abstracts and the uncertainty stemming solely from the estimation algorithm used for approximating the LDA model is presented in Section~\ref{application}. The concluding Section~\ref{conclusion} also provides some ideas for possible extensions and further applications of the tool.

\section{Measuring Uncertainty}
\label{measuring}

Uncertainty of estimators might result from different sources. For example, selecting different random samples from an underlying population of texts might change estimation outcomes as in any application of statistical methods. Furthermore, random components in the estimation algorithm such as random starting values or random sampling of candidate solutions result in stochastic outcomes of the estimation routine for a given fixed input. Besides such uncertainty related to stochastic components, the method introduced can also be applied, when the uncertainty is rather the result of deterministic decisions, e.g., with regard to choosing parameters of algorithms used in the pre-processing or modelling of the textual data. Then, the resulting figures do not represent uncertainty in the sense of a confidence interval, but rather the extent to which estimated topics are robust to changes in parameter values. 

For the illustration example in this paper, a real random component is considered, namely the one related to seeds of random number generators for starting values and updates in the estimation algorithm used to estimate a LDA model. Given a specific sample and keeping all other parameters fixed, the resulting topics should be identical if the algorithm is able to deliver an optimal solution with a finite, typically rather small, number of iterations. Thus, the interest is in the robustness of estimated topics for a limited number of iterations of the algorithm across restarts with different random seeds and how this might improve when the number of iterations is increased.

The general procedure, which is the same for repeated runs with stochastic components and for repeated runs with parameters which are modified in a deterministic way, comprises the following steps, which will be described in more detail below:
\begin{enumerate}
    \item Repeat the estimation procedure for all different settings.
    \item Match the resulting topics from different repetitions.
    \item Choose a reference set of topics.
    \item Calculate the distribution of word weights for a topic across repetitions as input for the graphical presentation described in Section~\ref{graphical}.
\end{enumerate} 

First, the estimation procedure has to be repeated for all different settings. Unfortunately, given the large size of textual data and the high complexity of estimation algorithms, this might result in high computational cost. However, given current available computing infrastructure, such analyses are feasible at least for exemplary cases.

\begin{figure}[H]
\centering
\vcenteredhbox{\includegraphics[width=0.45\textwidth]{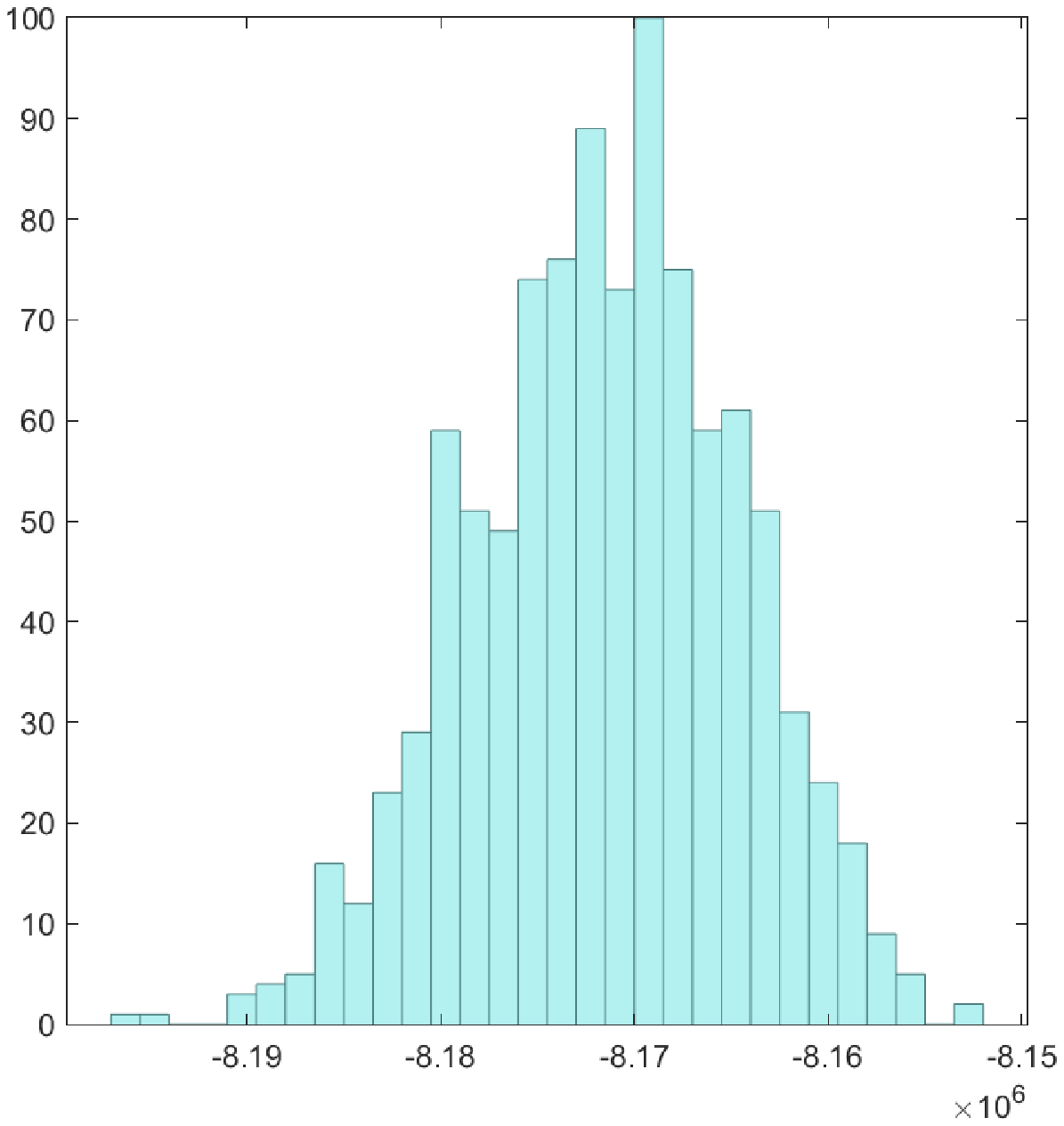}}
\vcenteredhbox{\includegraphics[width=0.45\textwidth]{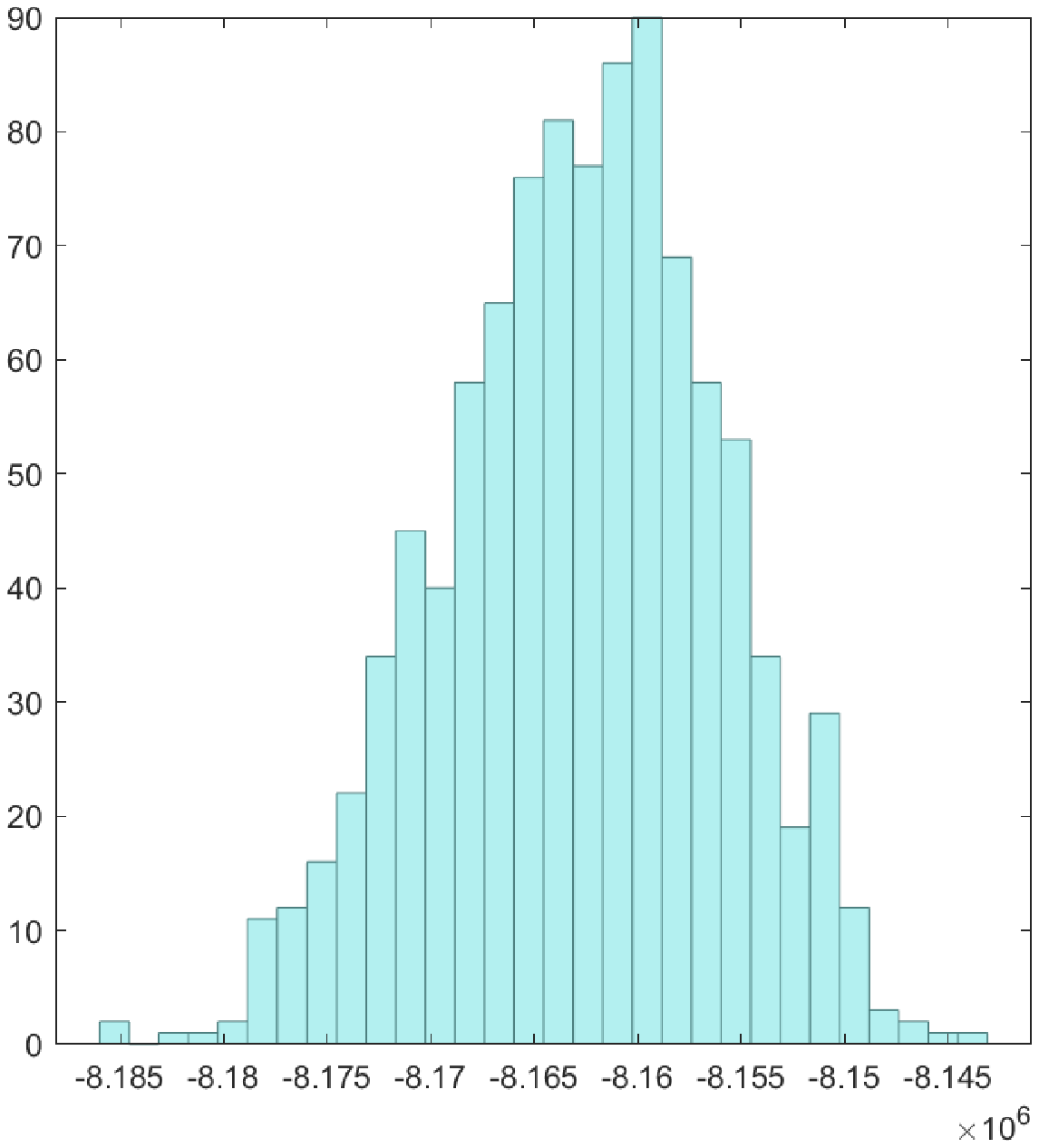}}
\caption{Histograms of Scores for 1\,000 Random Restarts with 10 (left) and 100 (right) Iterations.}
\label{likelihoods}
\end{figure}

As an example, Figure~\ref{likelihoods} shows the distributions of the approximate log-likelihood reported when estimating a LDA model to the same text corpus keeping all pre-processing steps and parameter values fixed, and just varying the random seed of the optimization routine. The left hand plot shows the distribution of 1\,000 restarts for a number of 10 iterations in the optimization algorithm, while the right hand plot shows the corresponding distribution when using 100 iterations.

It is obvious that there is substantial variation in the estimation results, which is solely due to the random seed for the stochastic optimization routine. Furthermore, it can be seen that this variation is reduced when spending more computational resources on the optimization procedure. However, considering only the results regarding the fit of the model is not sufficient for assessing the robustness of the estimation outcomes, which, in the case of a topic model, are the topics. Even small changes of the fit might correspond to large changes of the topics or, the other way round, quite similar topics might come with large differences in the model fit. Therefore, the eva

Second, a part from the sheer computational complexity of obtaining the estimation results, there is a second issue in topic modelling, namely that the topics are not identified in a statistical sense. At least the ordering of the topics resulting from an estimation procedure has to be considered as random. Therefore, in a second step, topics obtained in different simulation runs (e.g., different bootstrapped data samples, different random seeds for optimization routine, different parameter values) have to be matched, before a robustness analysis becomes sensible.

For the matching of topics from two estimates, the method proposed by \citeasnoun{Bystrov2022a} is applied. It consists in calculating the matrix of pairwise distances between the topic vectors from the different estimates. Thereby, different distance metrics might be used such as the cosine-similarity used in this paper or the Kullback-Leibler divergence. If the number of topics in both sets is the same as in the present application focusing solely on the effect of random restarts of the optimization algorithm, a one-to-one matching appears sensible. This means, that each topic in one of the sets is matched with exactly one topic in the other set. An efficient method for minimizing the sum of differences between the matched pairs for a one-to-one matching is the Hungarian algorithm \cite{Kuhn55}.

Third, if there exists one reference solution, e.g., when the corpora are generated by Monte Carlo simulation from a known LDA model as in \citeasnoun{Bystrov2022b}, a pairwise matching with this reference solution allows to identify all topics for each replication making use of the Hungarian algorithm. The problem becomes a bit more complex, if there is no known benchmark, e.g., in a setting when LDA models are fitted using different random seeds as for the application in this paper. In such settings, it is proposed to consider each of the outcomes as a potential benchmark solution and to perform a pairwise matching of all other outcomes to this pseudo-benchmark. Then, the pseudo-benchmark resulting in the overall smallest value of the loss function is selected for identifying the topic matches.\footnote{Alternatively to minimizing the sum of distances for all pairs, minimizing the maximum loss might be an option for future research.}

\begin{figure}[H]
\centering
\vcenteredhbox{\includegraphics[width=0.30\textwidth]{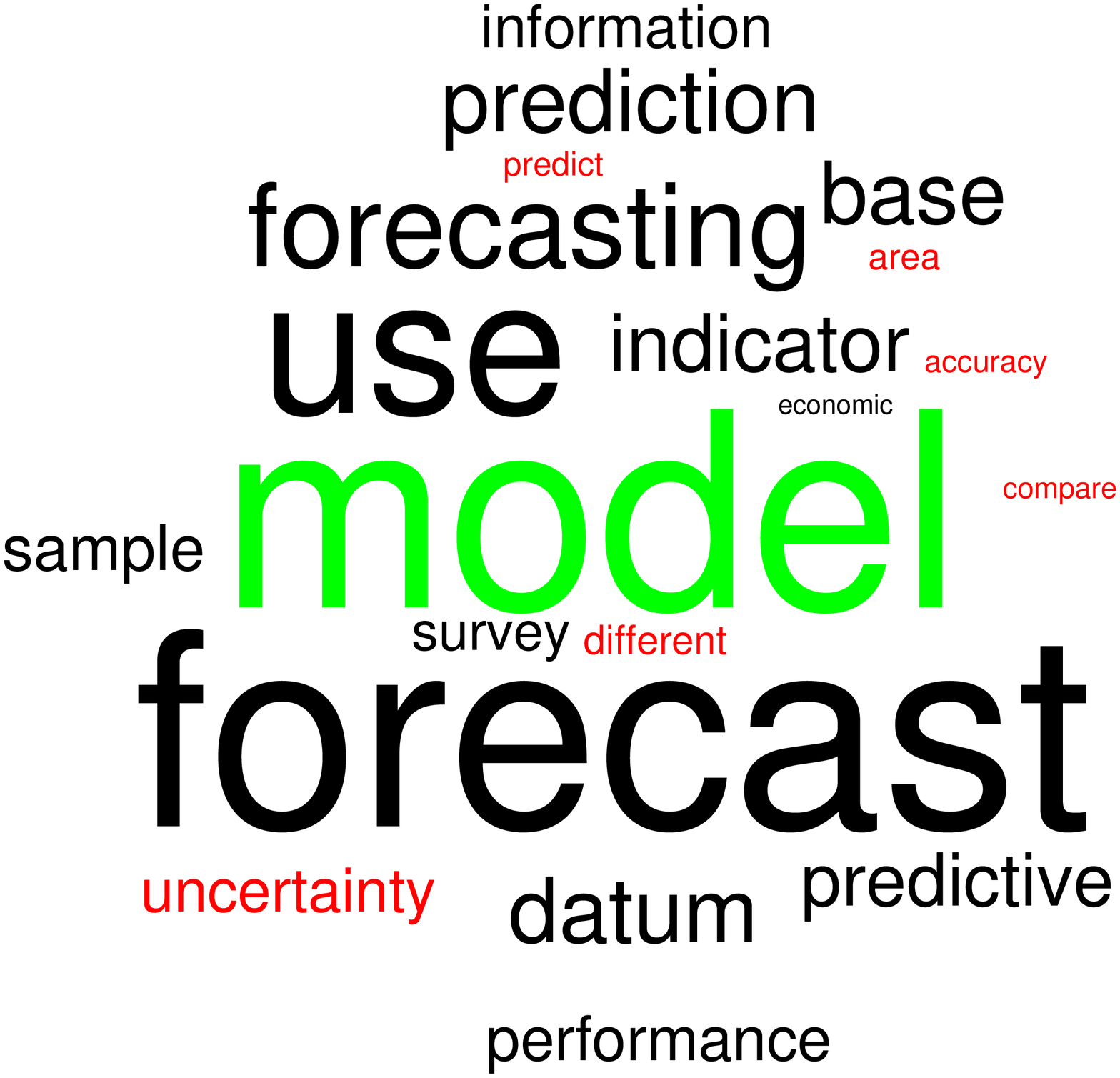}}
\vcenteredhbox{\includegraphics[width=0.30\textwidth]{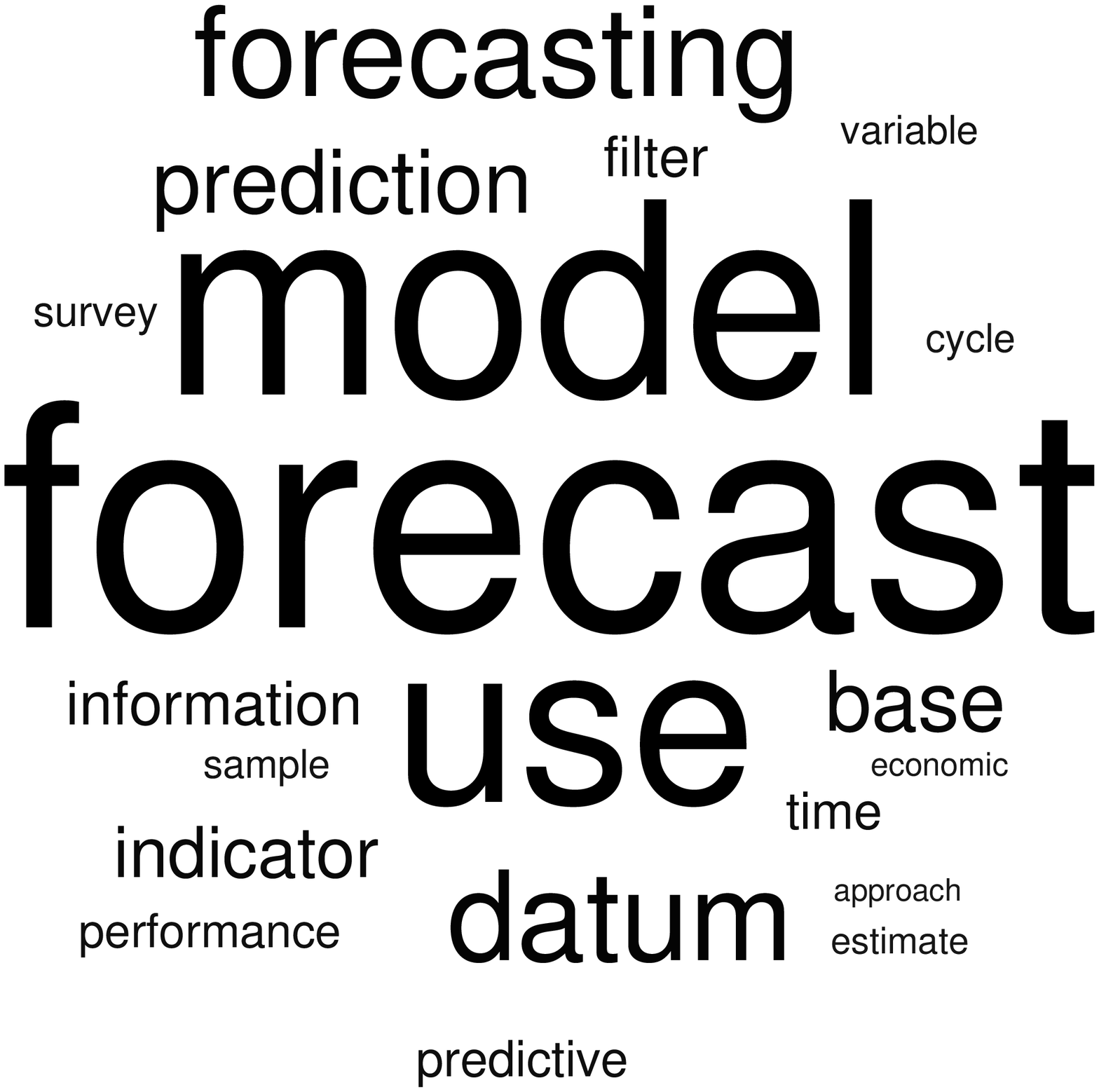}}
\vcenteredhbox{\includegraphics[width=0.30\textwidth]{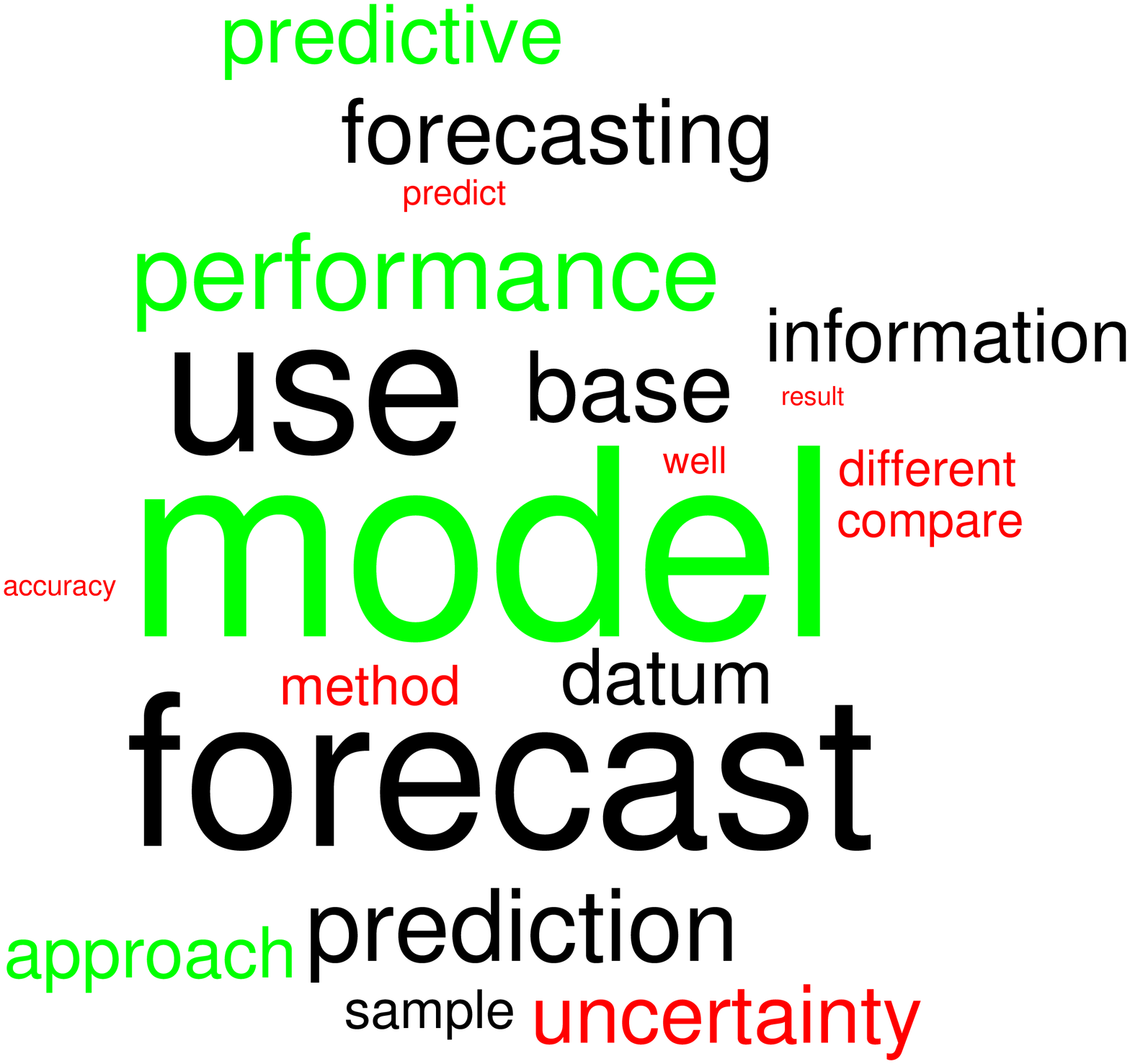}}
\caption{Similarity of Word Clouds for Different Random Seeds in LDA Model.}
\label{matched_topics}
\end{figure}

Figure~\ref{matched_topics} illustrates the results of this matching procedure for the application described in more detail in Section~\ref{application}. It is based on the matching of 1\,000 replications of the estimation of a LDA model with different random seeds with 100 iterations. It exhibits solely one word cloud from the reference model in the middle and two matched topics from two other estimates. Thereby, the word cloud on the left hand side corresponds to the estimate resulting in the lowest distance (highest cosine-similarity) to the reference topic after matching among all 999 other replications, while the right hand side shows the word cloud related to the match with the highest distance (lowest cosine-similarity) among the remaining replications. Thereby, for the two word clouds at the right and left of the reference word clouds, words in black colour correspond to words with weights differing by less than 25\% from those of the reference word cloud. Words in green colour indicate words showing up also in the top words of the reference word cloud, but with differences in weights exceeding 25\%, while words included among the top 20 words in the word clouds, which are not among the 20 words with highest weights in the reference word cloud, are shown in red.

The comparison shows that the differences in cosine-similarities also translate to visible differences in the distribution of top words in the word cloud. Nevertheless, it also becomes obvious that even for the worst match shown in the right hand side of Figure~\ref{matched_topics}, the topic remains visible for a human reader. 

Fourth, once, all topics have been identified across all replications, for each topic, the distribution of word weights across the simulated replications can be obtained, which is the necessary input for the graphical presentation described in the next section. Thereby, the term ``distribution'' does not correspond necessarily to a distribution in the statistical sense, such as for random seeds, but could also be just a collection of the outcomes for different parameter values.

\section{Graphical Presentation}
\label{graphical}

Given estimates for the uncertainty of word weights within topics, e.g., by empirical percentiles as described in the previous section, a final task consists in presenting these estimates in an accessible way. As mentioned before, a similar problem comes up when generating confidence sets or bands for time series forecasts or impulse response functions in multivariate time series models. In such applications, instead of providing long tables with the actual numbers, graphical presentations became a standard such as the fan-chart used by the Bank of England since 1996 \cite{Britton1998} for expressing the uncertainty related to forecasts. In this case, the object of interest is a plotted time series of forecast values for some periods ahead, which is surrounded by bands characterizing the extent of uncertainty. 

In direct analogy, for a word cloud, the objects of interests are the words, which again should be surrounded by {\it confidence bands} reflecting the uncertainty related to the estimation of word weights. This goal is achieved by superimposing the same word in different colours and sizes reflecting specific percentiles of the distribution obtained in the previous step. It should be noted that as presented here, only point-wise measures of uncertainty are provided, i.e., focusing on single words and not taking into account the dependence of uncertainty for several words.\footnote{For methods suitable for a joint modelling in the time series context see, e.g., \citeasnoun{LUTKEPOHL2020}. It has to be left to future research to develop comparable methods for the word cloud context.}

\begin{figure}[H]
\centering
\vcenteredhbox{\includegraphics[width=0.45\textwidth]{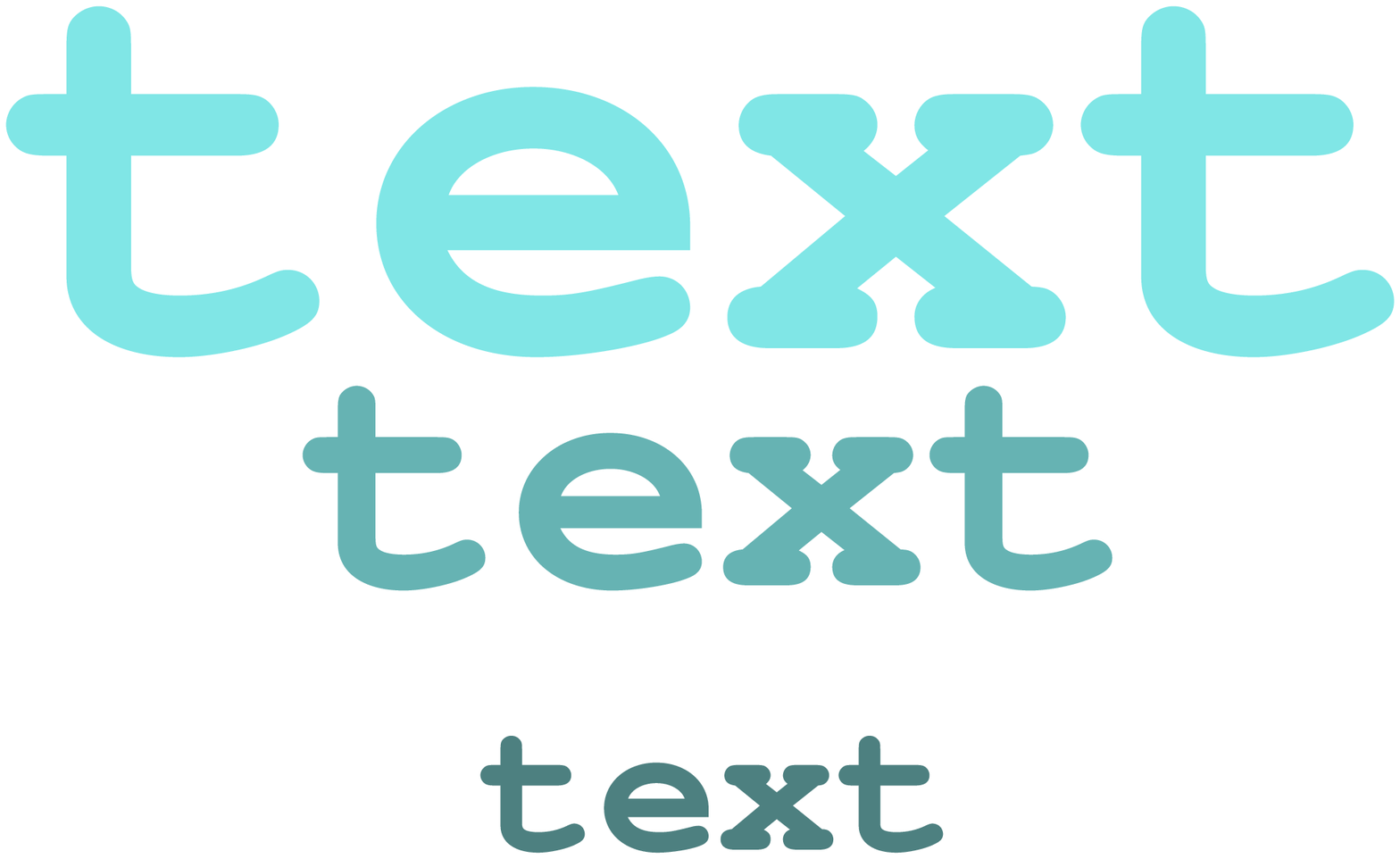}}
\vcenteredhbox{\includegraphics[width=0.45\textwidth]{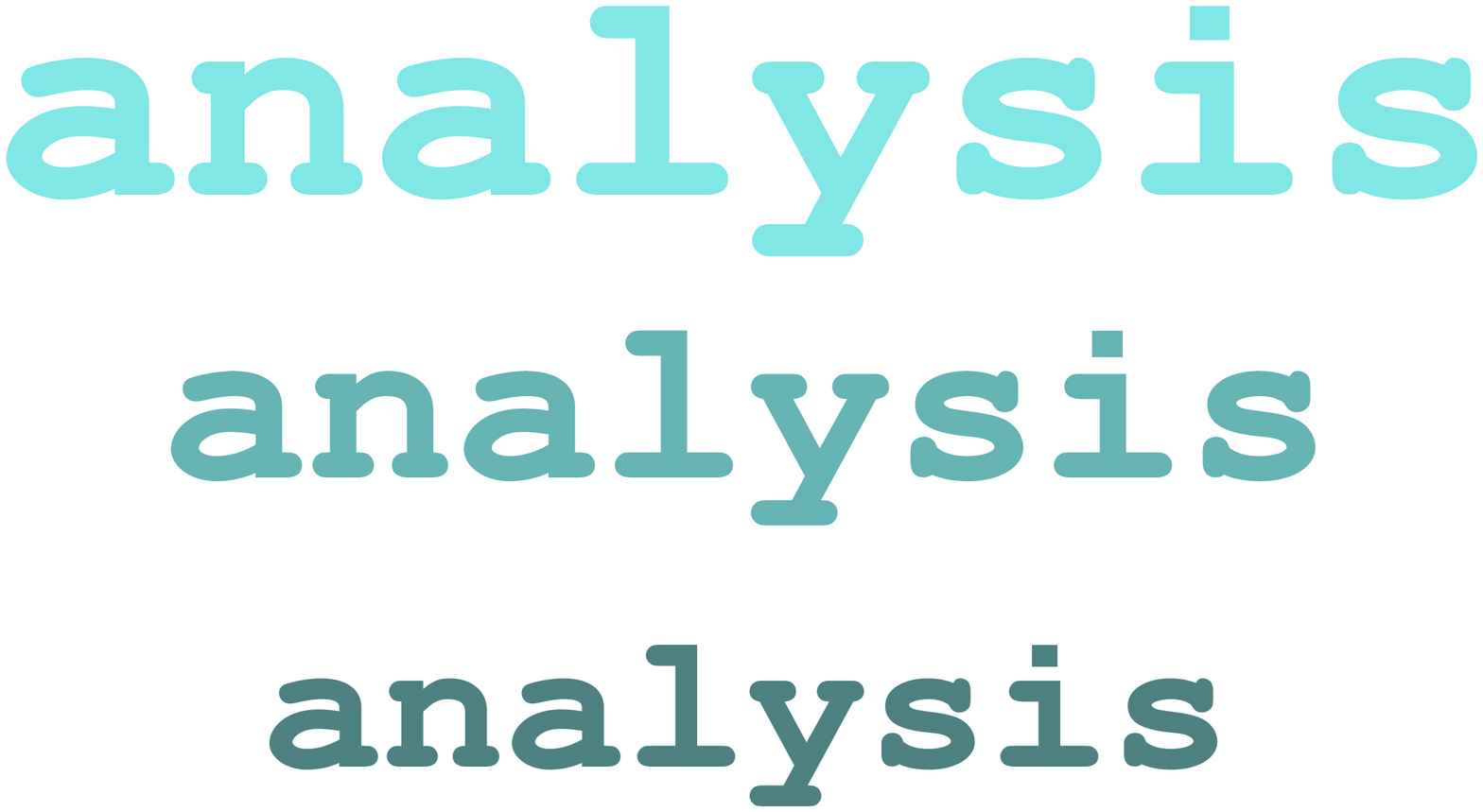}}
\caption{Examples of Words with Different Weights.}
\label{superimposed_words_1}
\end{figure}

Figure~\ref{superimposed_words_1} presents the first step of this approach, namely the presentation of the same word in different colours and sizes. From the analysis described in the previous section, for each word in a word cloud such as two words {\tt text} and {\tt analysis} used as examples here, weights corresponding to different percentiles are provided, e.g., the 10\%, 50\% and 90\% percentile. The font size of the three copies of both words shown in the left and right part of the figure, respectively, are proportional to the values of these percentiles. For the example, the 50\% percentiles of both words are set to the same value, while the surrounding 10\% and 90\% percentiles are closer to the 50\% percentile for {\tt analysis}.\footnote{The values used for this synthetic example are $(80,140,240)$ for {\tt text} and $(116,140,180)$ for {\tt analysis}.} In order to maintain a comparability also across words of different length as {\tt text} and {\tt analysis}, a monospaced font has to be used and the font size has to be adjusted for word length, i.e., a longer word with the same weight as {\tt text} will be presented in a smaller font size such that the areas of the rectangular box comprising the words are identical.\footnote{For the example, this implies that for the same weight, the font size of {\tt analysis} is rescaled by the factor $\sqrt{4/8}$ compared to the one of {\tt text}.}  

The smallest copy of the word {\tt text} on the bottom in the left part of the figure corresponds to the weight of the word at the 10\% percentile of distribution, the copy in the middle to the median, and the largest one on top to the 90\% percentile. The quite different sizes reflect a substantial amount of uncertainty for this word in the specific topic. The right part shows that the 10\% (bottom) and 90\% (top) percentiles are closer to the median for the word {\tt analysis}. It should be taken into account that the weights corresponding to the median are assumed to be the same for both words. Therefore, the area of a rectangle comprising the two words in the middle should be of the same size. 

It is worth noting that weights at the different percentiles can become very small such that the corresponding copies of the word become too small for presentation in a standard screen resolution. Furthermore, also the weights corresponding to the median might differ quite substantially for words within a topic. Therefore, word clouds typically only present the words with highest weights. The risk of percentiles becoming to small for graphical presentation increases with a decreasing median weight of a word.

\begin{figure}[H]
\centering
\vcenteredhbox{\includegraphics[width=0.45\textwidth]{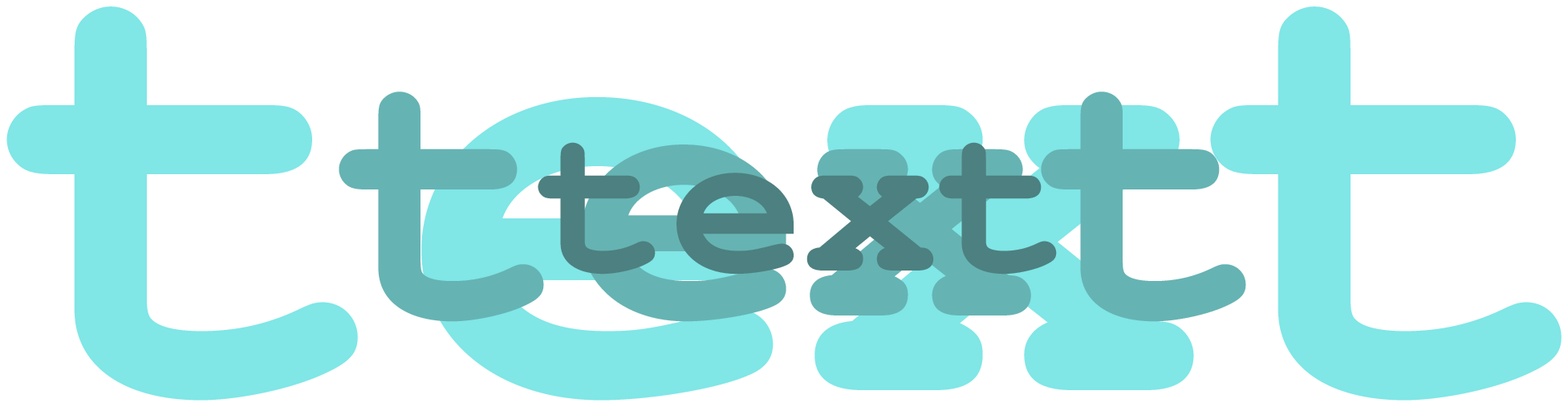}}
\vcenteredhbox{\includegraphics[width=0.45\textwidth]{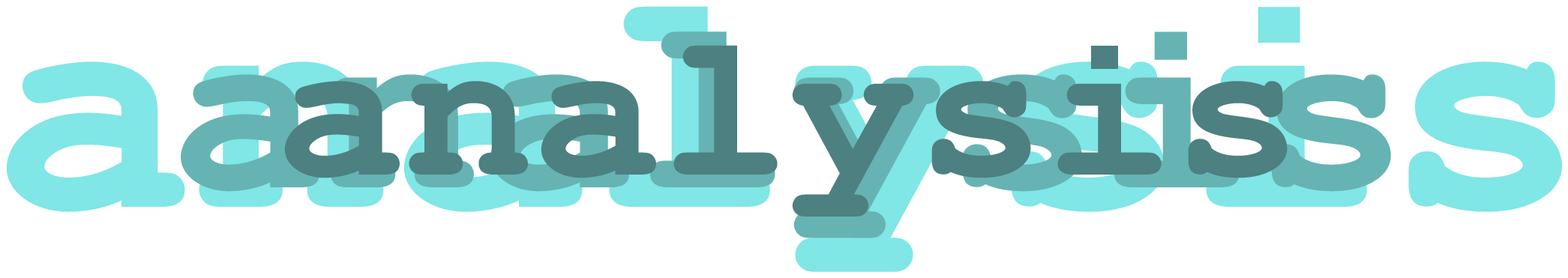}}
\caption{Examples of Superimposed Words.}
\label{superimposed_words_2}
\end{figure}

Figure~\ref{superimposed_words_2} presents the same three copies of the words as in Figure~\ref{superimposed_words_1}. However, now, the words are superimposed over the joint center of all bounding boxes. These representations should reflect the extent of the uncertainty due to the factor(s) considered in the analysis. If this uncertainty is large, the three copies might differ substantially in size such as the example {\tt test} on the left part of the figure. As mentioned above, quite uncertain estimates might include extreme cases, when some of the smaller percentiles are not visible anymore as their size falls below the limit set by screen resolution. Thus, if some words in a word cloud do not show all colours, this is evidence for large uncertainty. On the other side of the plot, the three copies of {\tt analysis} differ less in size, which points towards a more robust finding for the specific word in the specific topic, i.e., a smaller ``confidence intervall''.

While the outline of the method of presenting ``confidence intervals'' for words within word clouds is provided for only three percentiles, it generalizes easily to five or more percentiles. However, the reader's ability to differentiate sizes and colours puts a natural limit to the number of percentiles provided. Furthermore, it should be noted that the results for a single word should not be re-scaled after the procedure in order to maintain the size ratio between words (even across topics if such a comparison is of interest). Only uniform scaling for all percentiles of all words in all word clouds will not affect the interpretation of the resulting word clouds with ``confidence intervals'' for the words. 

The final step of the procedure is organizing the superimposed words belonging to a word cloud in a figure, i.e., the $k$ words for a topic exhibiting the highest weights for that topic. While there exist very efficient toolboxes for this purpose, e.g., {\tt word\_cloud} in Python or {\tt wordcloud()} in Matlab, they cannot be used without further adjustments for presenting word clouds including the type of ``confidence intervals'' described above. The Python toolbox {\tt word\_cloud} re-scales font sizes to obtain a good fit of the words into a predefined shape, and for the Matlab the source code is not available, such that there is no straightforward way to superimpose the words in different sizes. Therefore, I propose using a general purpose heuristics optimization algorithm, Threshold Accepting, for organizing the superimposed words into word clouds.\footnote{Obviously, it is rather straightforward on how to adjust existing algorithms for word clouds to allow for ``confidence intervals'' for words as suggested in this paper.}  Threshold Accepting was introduced by \citeasnoun{Dueck1990} as an alternative to Simulated Annealing and has been used successfully for many problems also in statistics \cite{Winker2001}.

The task to be solved by the Threshold Accepting implementation is stated as follows: 

\begin{enumerate}
\item The rectangles comprising the largest copy of the superimposed words should not overlap.\footnote{Given the different height of letters, a more precise description of the envelope might result in more compact presentations. However, such adjustments are left for future implementations.}
\item Rectangles containing words with high weight in the topic should be closer to the center of the figure.
\item The overall canvas for the figure is a square. All words have to fit on the canvas (binding constraint), but a form of the word cloud closer to a circle is preferred (non-binding constraint). 
\end{enumerate}

To achieve this goal, in each search step of the algorithm up to two words can make a small random move, two words might change place or one word might rotate by 90 degrees. Thereby, an objective function including the three goals with different weights is to be minimized. As a local search heuristic, Threshold Accepting can evade sub-optimal local solutions. It is important to note that it is not necessary to obtain a global optimal solution for this complex knapsack problem. It is enough to find a solution which satisfies the binding constraints and achieves sufficient quality for the other constraints. However, the number of local search steps of the Thresholds Accepting implementation has to be quite last to obtain satisfying solutions. Therefore, future analysis might consider using standard algorithms for generating word clouds and how to adjust them for the setting with ``confidence intervalls'' around words. 

\begin{figure}[H]
\centering
\vcenteredhbox{\includegraphics[width=0.9\textwidth]{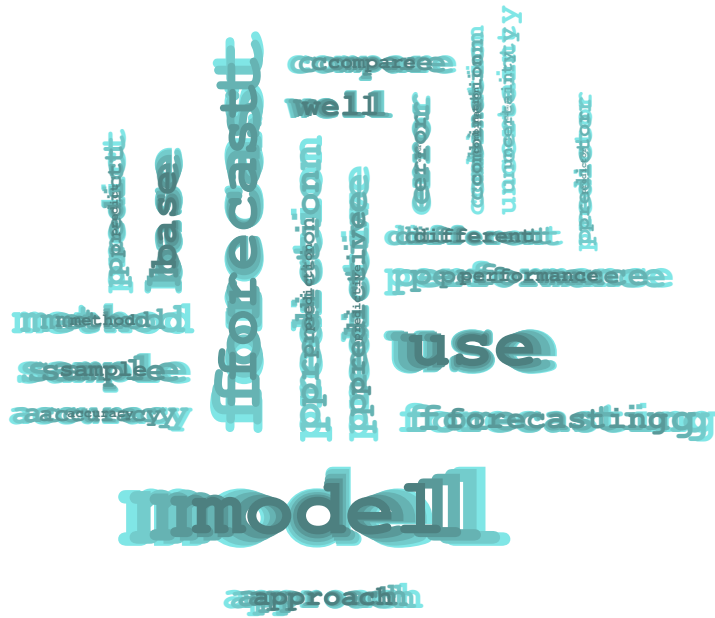}}
\vspace*{-2cm}

\caption{Example of Word Cloud with ``Confidence Intervals'' for Words.}
\label{example_wordcloud}
\end{figure}

An example of a world cloud of the 20 most important words from a topic linked to the application discussed in the following section is provided in Figure~\ref{example_wordcloud}. It exhibits the 10, 20, 50, 80, and 90 percent quantiles of the distribution from 100 restarts of the LDA estimation algorithm. 

As for a standard word cloud, one immediately sees that {\tt model} and {\tt forecast} are the words with the highest weight in this topic. And for these words also the ``confidence intervals'' are not very large as all five superimposed copies of the words are of similar size. This holds true for most of the words in the word cloud, but there are also exceptions as the word {\tt uncertainty} in the top right part, where the differences are substantially larger. The 20\% percentile is still visible, but very small, while the 10\% percentile is too close to zero for a graphical presentation. Thus, uncertainty about the membership of this word to the topic stemming solely from the optimization algorithm is substantial. Nevertheless, overall one could conclude that this topic is rather robust regarding repeated applications of the LDA estimation algorithm. 

\section{Application to Conference Abstracts}
\label{application}

The performance of the new method for visualizing the uncertainty surrounding word clouds is illustrated using the topics of abstracts from a conference series in statistics covering the period 2007 to 2021. The eventual goal of the analysis is identifying topics which describe well the contributions to this conference series and might be used to track changes in scientific content over time. Here, the focus is solely on the robustness of the identified topics with regard to uncertainty stemming from the random component of the used estimator.

The conference series considered is the International Conference on Computational and Financial Econometrics (CFE) which is run jointly with the International Conference of the ERCIM WG on Computational and Methodological Statistics (CMStatistics).\footnote{See, e.g., \url{http://www.cmstatistics.org/CMStatistics2022/}.}

\begin{figure}[H]
\centering
\vcenteredhbox{\includegraphics[width=0.9\textwidth]{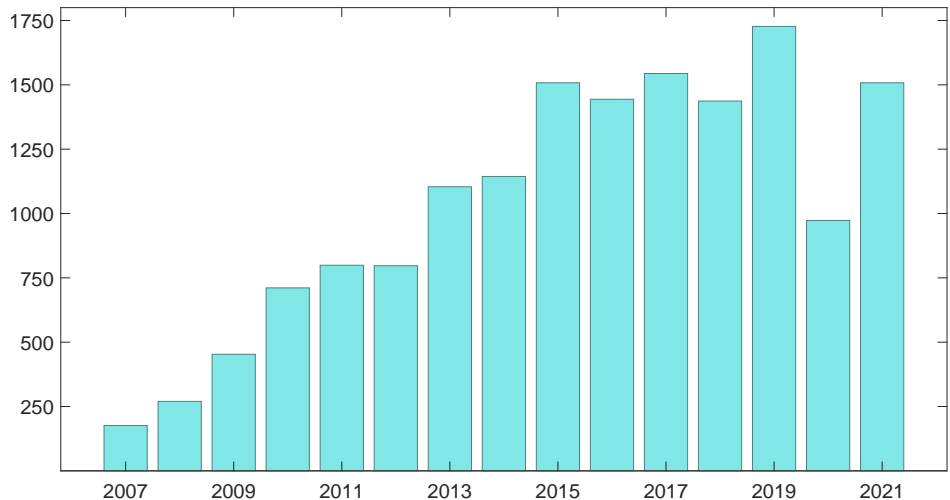}}

\caption{Number of Abstracts in CFE-CMStatistics Conferences 2007 -- 2021.}
\label{nofabstracts}
\end{figure}

Figure~\ref{nofabstracts} shows the development of the number of abstracts published in the book of abstracts of these conferences during the period 2007 -- 2021. It indicates that the conference series grew from a small scale meeting within a few years to one of the largest conferences in the domain of computational statistics and econometrics in Europe. Obviously, the number of more than 15\,500 abstracts precludes a manual analysis of topics and their importance for the conference.

The book of abstracts for this conference series exhibits a clear structure, which changed only few times in the course of the years. Therefore, it was possible to extract the individual abstracts including metadata on authors, title and country of institutional affiliation in an automated process. The abstracts are not complemented by keywords or some code for specific areas such as JEL-codes or AMS-classes. After extraction and splitting, all abstracts were pre-processed by removing very common words (stop words), punctuation, numbers and words which appear too rarely or too often to be informative about specific topics. While all these pre-processing steps might have an impact on the identified topics, these effects are not studied in further detail as the focus of the illustrative application is on the effect of random seeds in the estimation algorithm and the visualization of these effects. 

A further important decision in topic modelling is about the number of topics. Different methods are used for this purpose, which are not subject of the present analysis.\footnote{For a recent comparison of some alternatives in a Monte Carlo study including a new information criterion see \citeasnoun{Bystrov2022b}.} Instead, the number is fixed to 20, which was found in preliminary analysis based on subjective evaluation of the results to lead to sensible topics. 

It is also not the aim of this illustration to study the impact of other parameters of the estimation routine apart from the random seeds. Therefore, a standard implementation of LDA in Python is used with standard settings for all other parameter values.\footnote{The methods CountVectorizer and LatentDirichletAllocation from sklearn have been used. The cut-off values for word frequencies were set to 75\% and 0.5\% respectively. The learning method for LDA was set to `online' with a learning offset of 50. All other parameters were left at their default values.}  Only the number of iterations is varied in order to see whether more iterations contribute to reducing the uncertainty in estimated word clouds due to the randomness of the optimization algorithm. 

\begin{figure}[H]
\centering
\vcenteredhbox{\it \footnotesize Forecasting \hspace*{2cm} Financial Risk \hspace*{2cm} Extreme Value}
\vspace*{0.7cm}

\vcenteredhbox{\includegraphics[width=0.26\textwidth]{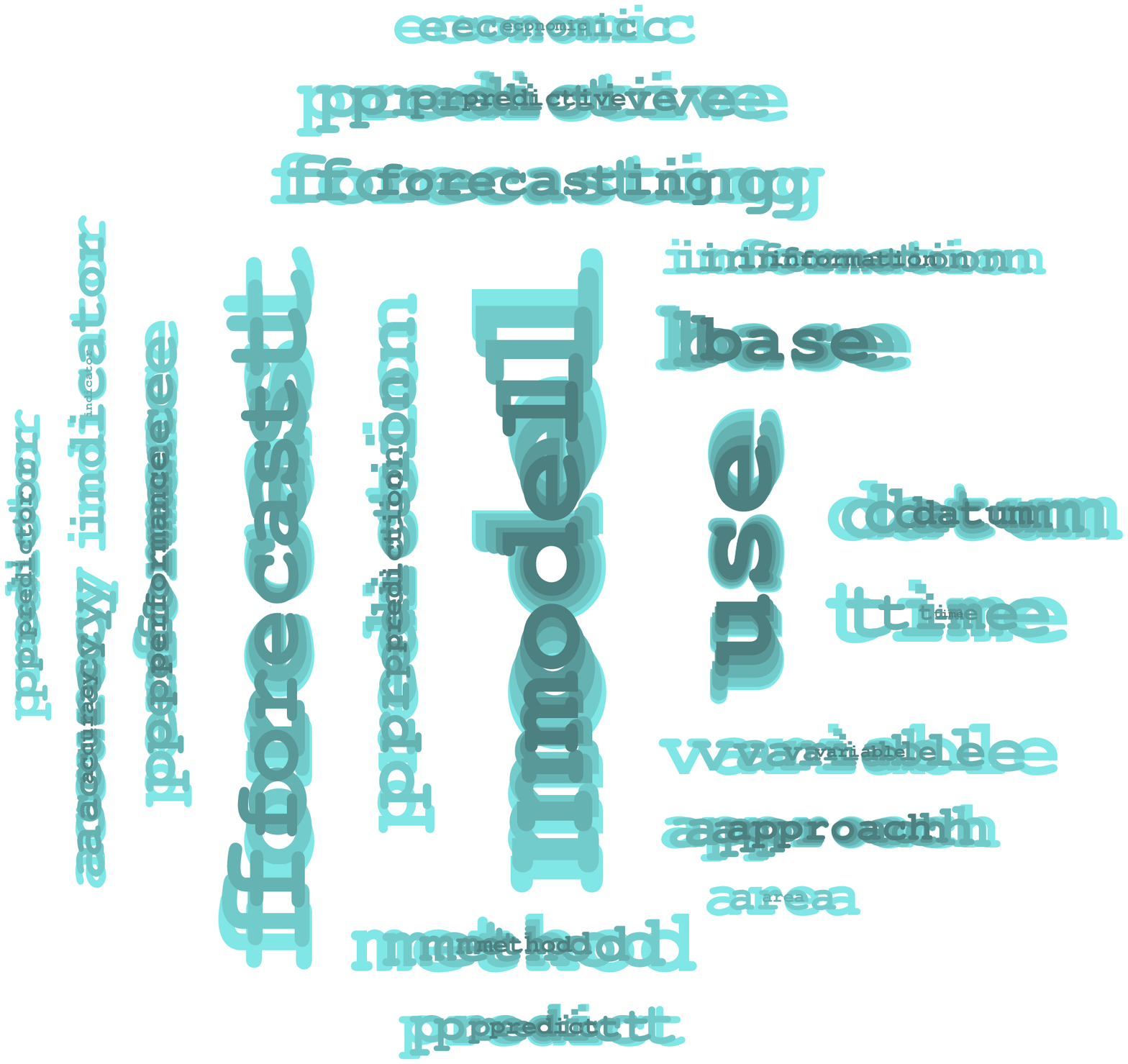}} \hspace*{0.5cm}
\vcenteredhbox{\includegraphics[width=0.26\textwidth]{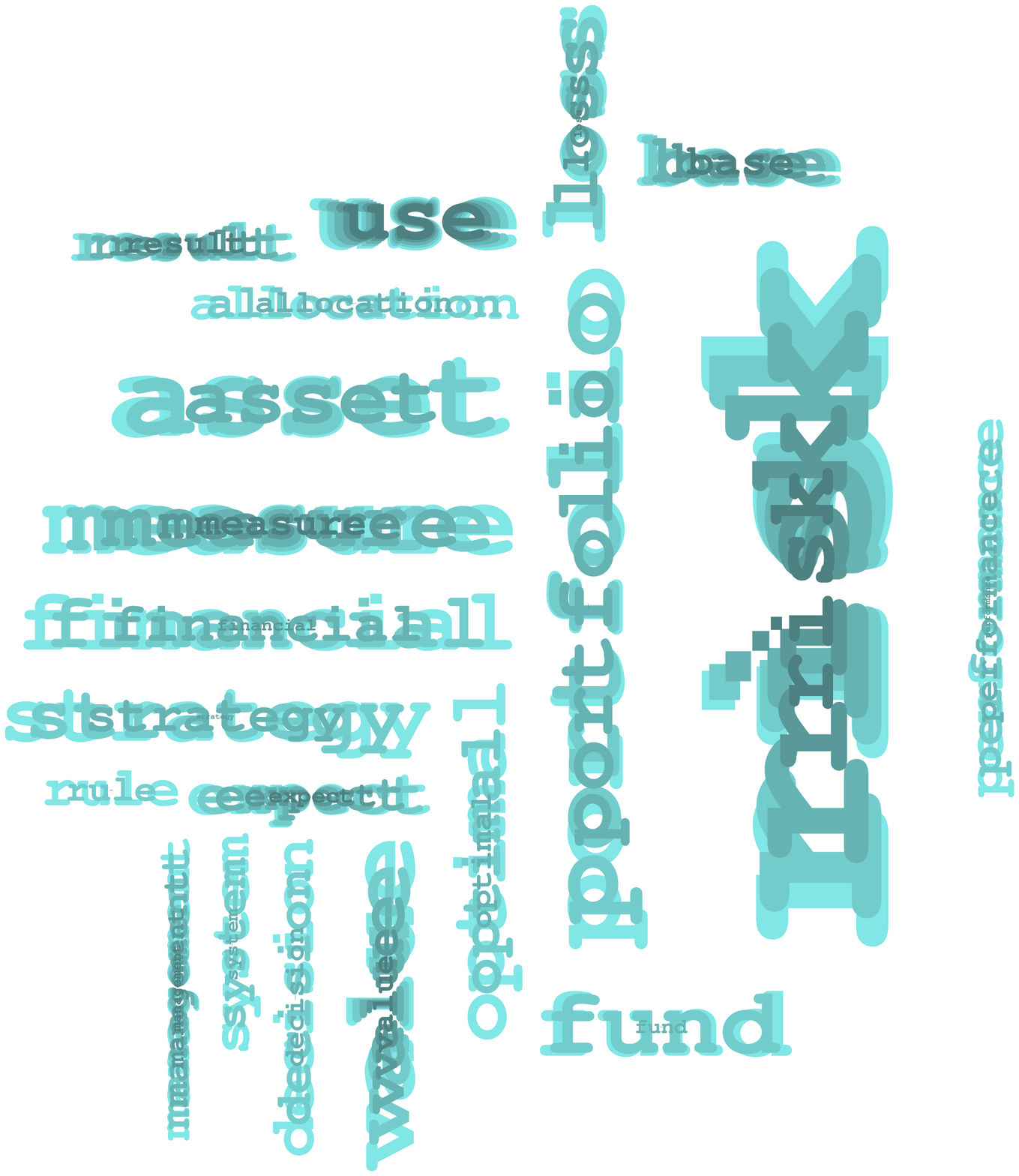}} \hspace*{0.5cm}
\vcenteredhbox{\includegraphics[width=0.26\textwidth]{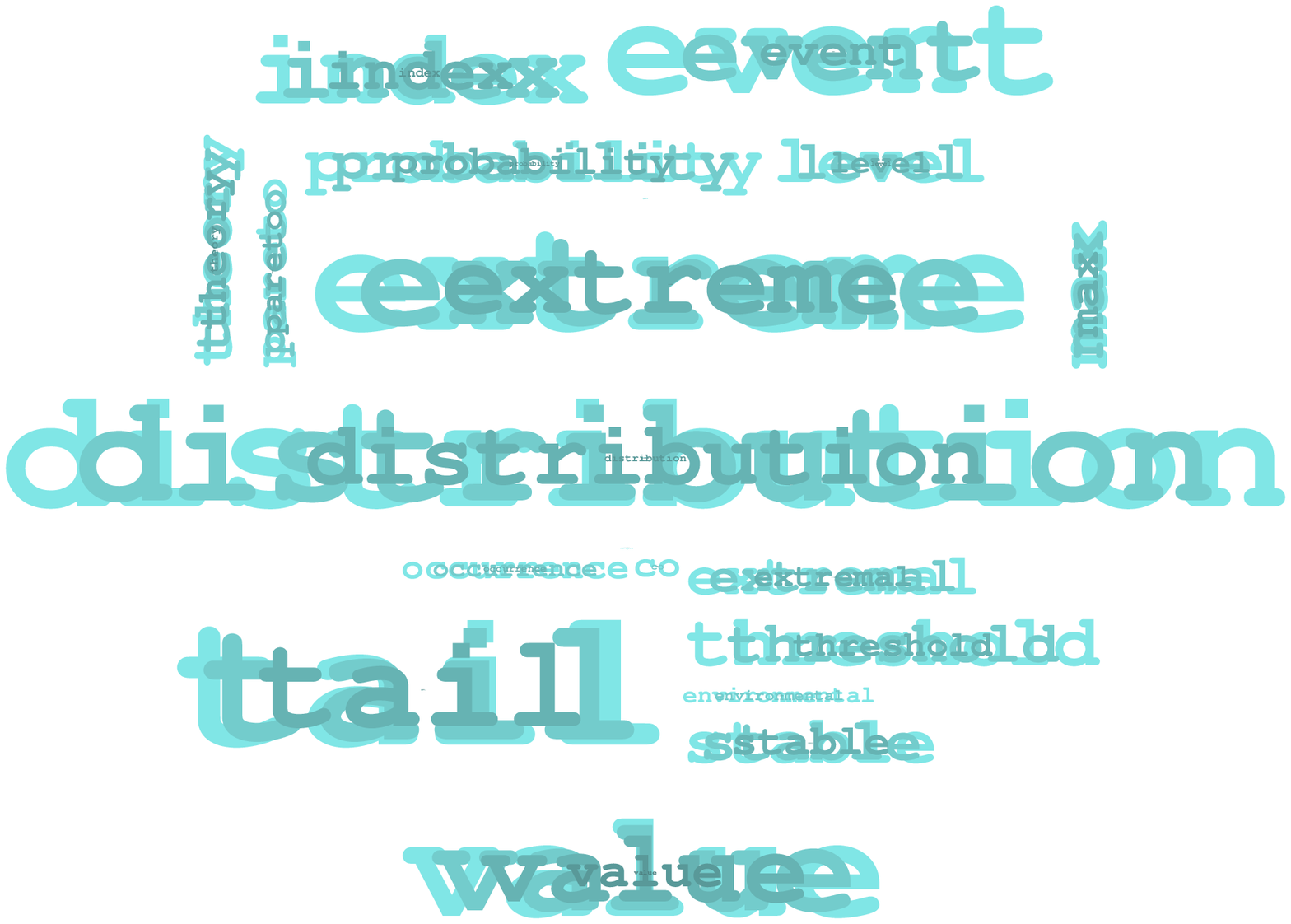}}
\caption{Word Clouds with ``Confidence Intervals'' for 10 Iterations.}
\label{examples_10}
\end{figure}

Figure~\ref{examples_10} shows three word clouds with their 10\%, 20\%, 50\%, 80\% and 90\% percentiles from the empirical distribution with 1\,000 replications of the estimation algorithm with 10 iterations.\footnote{It should be noted that finding the reference model out of the 1\,000 replications resulting in the smallest overall losses from matching is computational intensive as almost 1\,000\,000 matchings have to be calculated. In contrast, the threshold accepting algorithm used for constructing the word clouds requires only few seconds although about 1\,000\,000 search steps are conducted.} The first topic on {\it forecasting} appears to be quite well identified, while still exhibiting a substantial amount of variation of the word weights within the topic. The same holds true for the second topic on {\it Financial Risk}. The third word cloud related to {\it Extreme Value} appears to come with a much higher degree of uncertainty as for many of the top words only the weights of the higher percentiles shown in brighter colours are large enough to show up in the graph. 

\begin{figure}[H]
\centering
\vcenteredhbox{\it \footnotesize Forecasting \hspace*{2cm} Financial Risk \hspace*{2cm} Extreme Value}
\vspace*{0.7cm}

\vcenteredhbox{\includegraphics[width=0.26\textwidth]{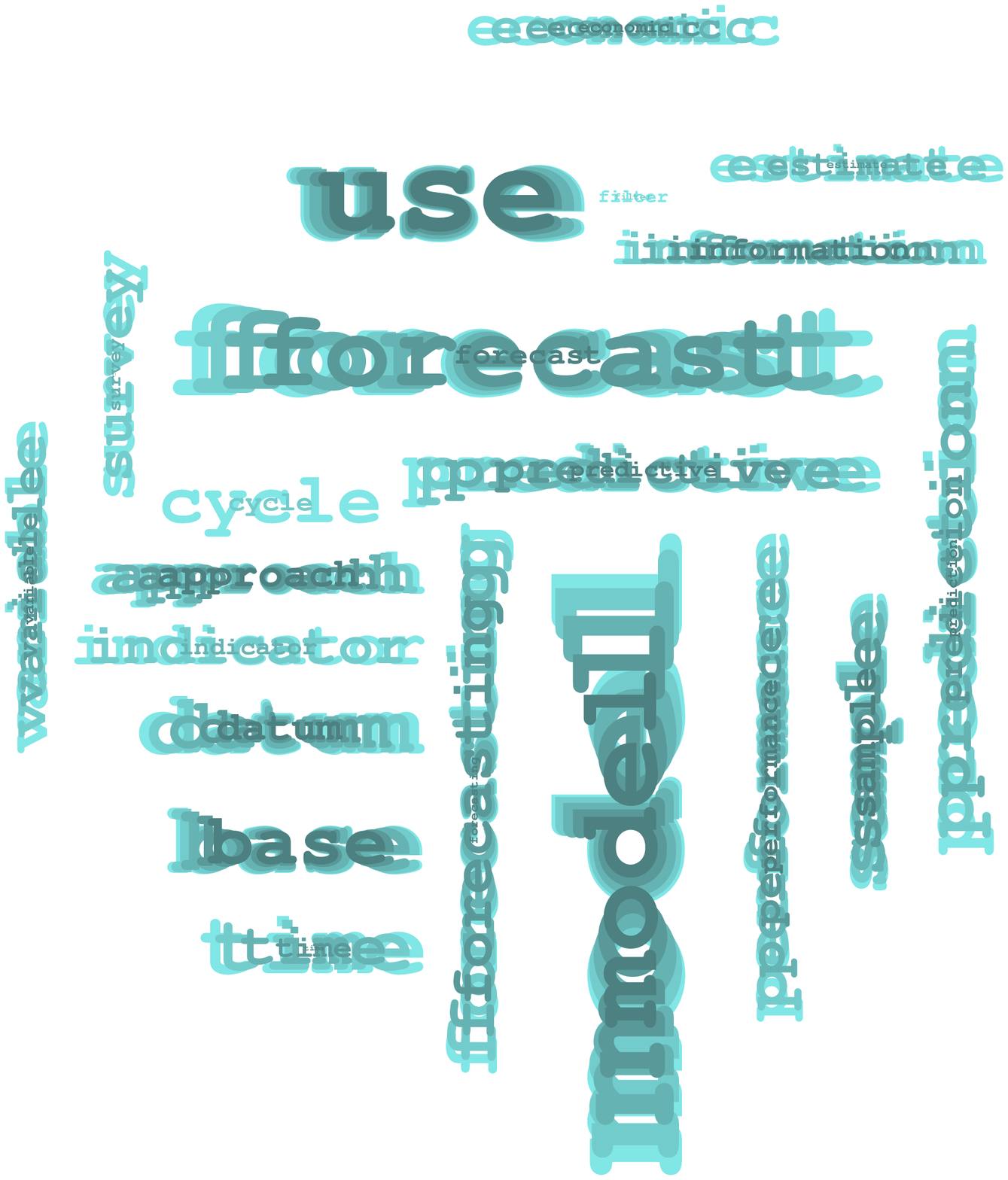}} \hspace*{0.5cm}
\vcenteredhbox{\includegraphics[width=0.26\textwidth]{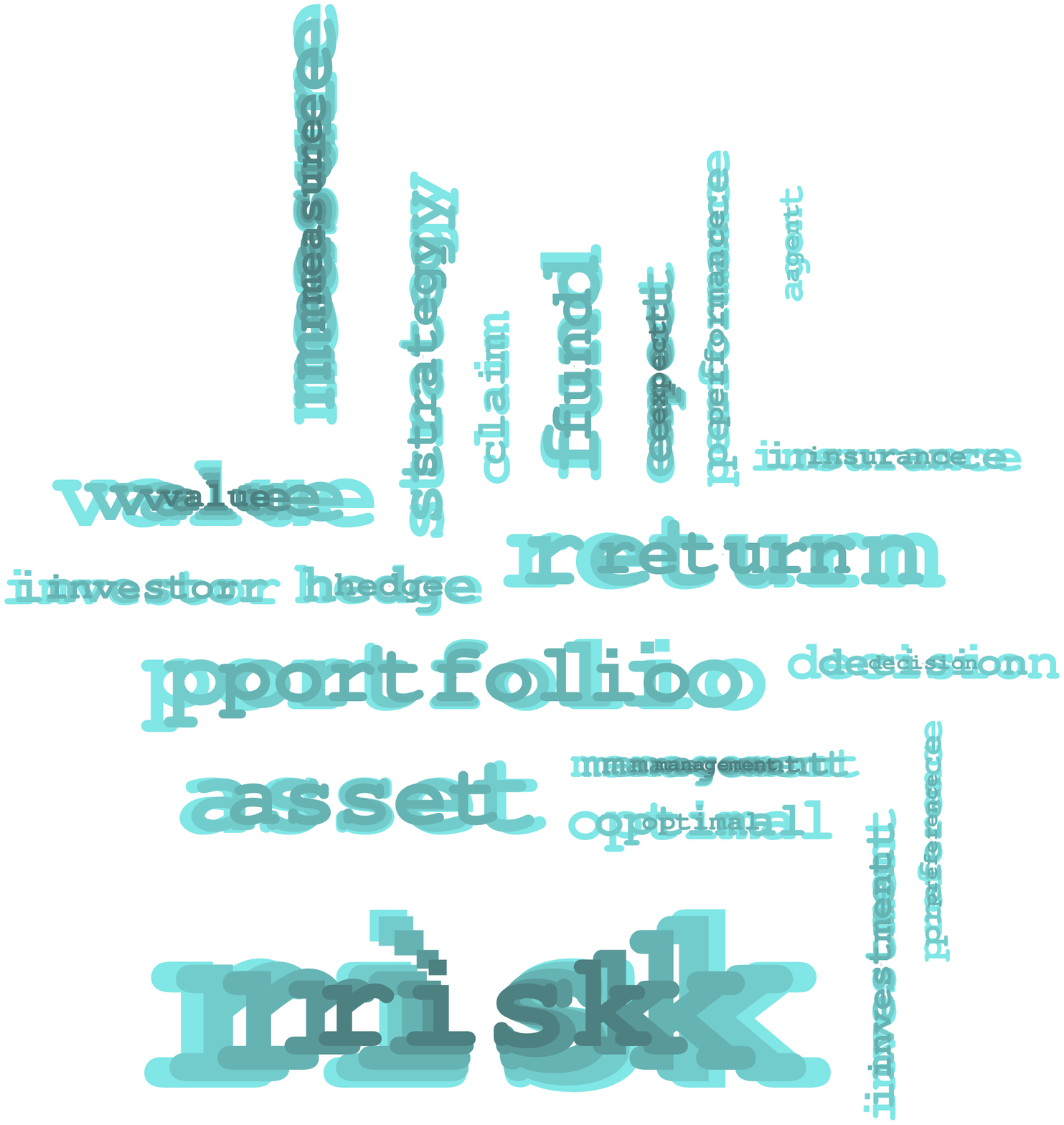}} \hspace*{0.5cm}
\vcenteredhbox{\includegraphics[width=0.26\textwidth]{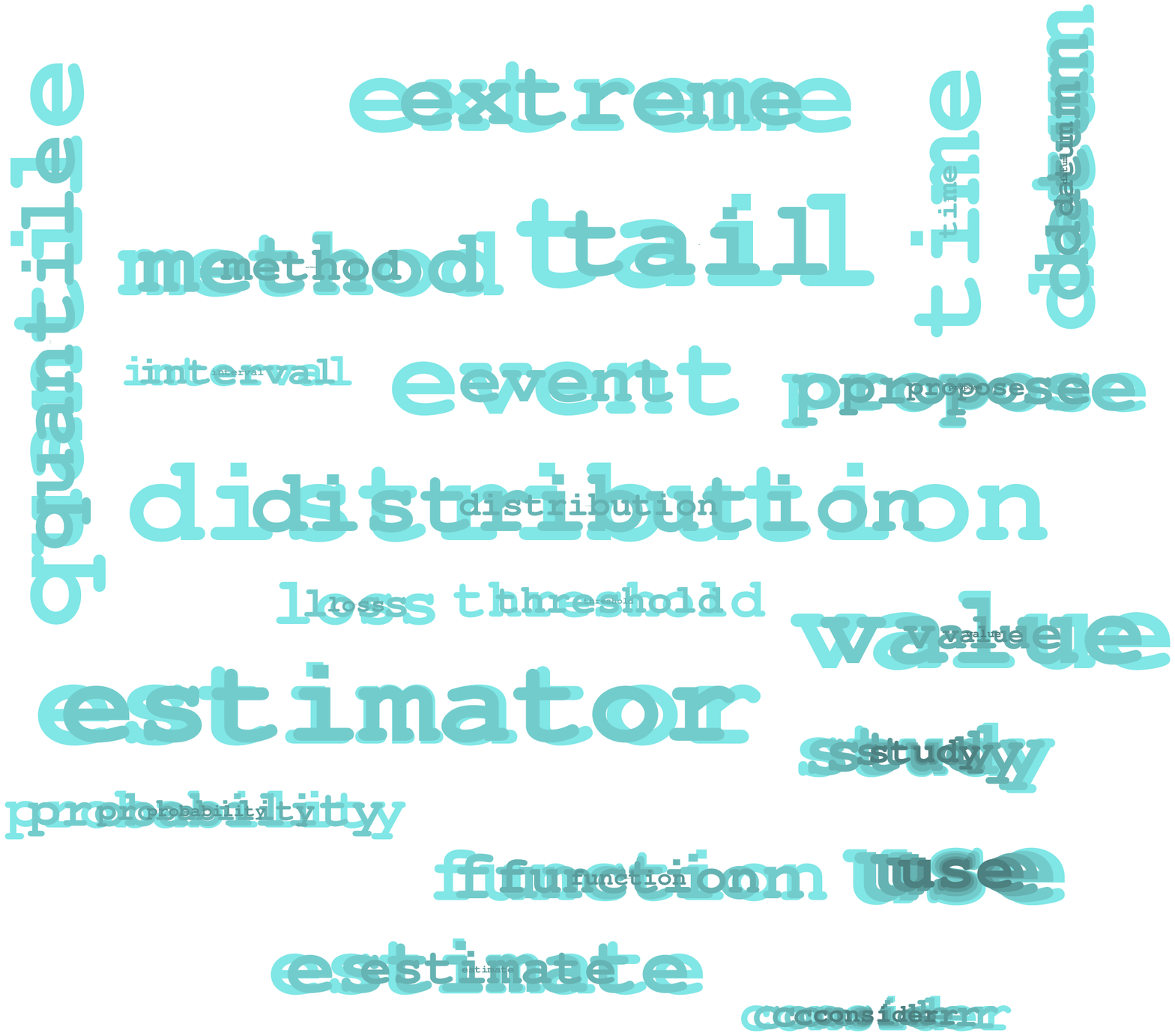}}
\caption{Word Clouds with ``Confidence Intervals'' for 100 Iterations.}
\label{examples_100}
\end{figure}

Given that the application example is using real data, the real data generating process is not known. Thus, it cannot be assessed to what extent the different quality of estimates of the word clouds is a result rather of a model miss-specification or of the randomness of the estimation procedure. However, by increasing the iterations for the estimation algorithm, the latter effect might be reduced. Thus, the analysis was repeated using 100 iterations within each of 1\,000 replications of the estimation procedure. Figure~\ref{examples_100} show the corresponding results for the same three topics. It should be noted that a new reference model was selected based on the 1\,000 replications with 100 iterations. Thus, it might not be surprising to see some new words like `cycle' in the first topic showing up with low weights.

Overall a comparison of the word clouds for the same topic obtained based on 10 and 100 iterations, respectively, shows that the estimates of the most important words appear to become a bit more precise for the topics already estimated quite well with a low number of iterations. For the third topic, which exhibited a large degree of uncertainty due to the random component of the algorithm, is not estimated more precisely with a larger number of iterations, though. Given that the true DGP is not known in this real application, it is not possible to interpret this finding in a qualitative sense. This will have to be done in a controlled Monte Carlo simulation setting, when the true DGP is known. As the purpose of this paper is to introduce the new visualization method, such a Monte Carlo simulation is left for future research.

\section{Conclusion and Outlook}
\label{conclusion}

A novel method for simulation based quantification of uncertainty in estimation of LDA models is presented. It takes into account the problem of missing identification of topic numbers using a matching algorithm. The resulting distributions of word weights within topics are visualized by superimposing the same words in different colours and font size according to selected percentiles of the empirical distribution. This tool allows for an intuitive assessment of the precision of word cloud estimates.

An application to conference abstracts of the CFE-CMStatistics conference series demonstrates the method with regard to the uncertainty resulting from the estimation algorithm for the LDA model. It could be seen, how increasing the number of iterations of the optimization algorithm for estimating the LDA model reduces the uncertainty of word clouds at least for word clouds which appear rather stable already for a small number of iterations. However, even with a rather large number of 100 iterations,\footnote{Preliminary results with 1\,000 iterations do not change this finding substantially.} still a substantial amount of uncertainty persists, which should be taken into account when using word or topic weights for further analysis. 

The method can be used for several other settings of high interest, e.g., studying the effect of sampling or parameter settings both in the pre-processing of documents and for the LDA model. Furthermore, the method might be assessed with more rigor on corpora obtained by simulation from a known DGP as done in \citeasnoun{Bystrov2022b} for evaluating the performance of model selection criteria in the LDA context. In any case, the method allows to communicate uncertainty about model estimates in an (almost) as intuitive way as presenting the word clouds themselves.

A future extension of the approach might consist in taking into account the dependence between the measures of uncertainty for different words within a topic. This might be done following related approaches in the context of joint confidence bands for impulse response functions in multivariate time series models \cite{LUTKEPOHL2020}.

\bibliography{Visualizing_literature}
    
\end{document}